\documentclass[]{interact}
\usepackage[utf8]{inputenc}
\usepackage[english]{babel}
\usepackage{amsmath}
\usepackage{amsfonts}
\usepackage{amssymb}
\usepackage{graphicx}

\usepackage[sort&compress,numbers]{natbib}
\usepackage{booktabs}
\usepackage{nowidow}

\usepackage[per-mode=reciprocal]{siunitx}
\usepackage{todonotes}

\newcommand{\orcid}[1]{} 

\theoremstyle{plain}

\theoremstyle{definition}

\theoremstyle{remark}

\usepackage{hyperref}
\usepackage[utf8]{inputenc}

\newcommand{\dx}[1]{\left.\text{d}#1 \right.}

\newcommand{\R}{\mathbb{R}}
\newcommand{\normal}{n}

\def\SymbReg{\textsuperscript{\textregistered}}

\newcommand{\subcool}{\text{cool}}
\newcommand{\subrad}{\text{rad}}
\newcommand{\subamb}{\text{amb}}
\newcommand{\subvessel}{\text{vessel}}
\newcommand{\subblood}{\text{b}}
\newcommand{\subnative}{n}
\newcommand{\subcoag}{c}

\newcommand{\subinit}{\text{init}}

\newcommand{\temperature}{T}
\newcommand{\radiation}{\varphi}
\newcommand{\damage}{\omega}

\newcommand{\conductivity}{\kappa}
\newcommand{\density}{\rho}
\newcommand{\heatcapacity}{C_p}
\newcommand{\perfusion}{\xi}

\newcommand{\absorption}[1]{\mu_{a#1}}
\newcommand{\scattering}[1]{\mu_{s#1}}
\newcommand{\anisotropy}[1]{g_{#1}}
\newcommand{\coolingfactor}[1]{\beta_{q#1}}
\newcommand{\timehorizon}{\tau}
\newcommand{\htc}{\alpha}
\newcommand{\diffusivity}{D}

\newcommand{\laseron}{t_\text{on}}

\newcommand{\frequencyfactor}{A}
\newcommand{\activationenergy}{E_a}
\newcommand{\gasconstant}{R}

\newcommand{\n}[1]{\mathbf{ #1}}

\newenvironment{eqsystem*}
{\begin{equation*}
	\left\lbrace
	\begin{alignedat}{2}
}
{\end{alignedat}
	\right.
	\end{equation*}
}

\begin{document}
\articletype{PREPRINT}

\title{Validating a Simulation Model for Laser-Induced Thermotherapy Using MR Thermometry}

\author{\name{Frank H\"ubner$^{a}$,
Sebastian Blauth$^{b}$\orcid{0000-0001-9173-0866},
Christian Leith\"auser$^{b}$\orcid{0000-0001-8936-9805},
Roland Schreiner$^{a}$,
Norbert Siedow$^{b}$\orcid{0000-0003-0522-7717},
Thomas J. Vogl$^{a}$}
\affil{$^{a}$Institute for Diagnostic and Interventional Radiology of the J.W. Goethe University Hospital, Germany\\$^{b}$Fraunhofer ITWM, Kaiserslautern, Germany}
}

\thanks{Email: \href{mailto:christian.leithaeuser@itwm.fraunhofer.de}{\nolinkurl{christian.leithaeuser@itwm.fraunhofer.de}}}

\maketitle

\begin{abstract}
Laser-induced interstitial thermotherapy (LITT) is applied to ex-vivo porcine livers. An artificial blood vessel is used to study the cooling effect of larger blood vessels in proximity to the ablation zone. The same setting is simulated using a model based on partial differential equations (PDEs) for temperature, radiation, and tissue damage. The simulated temperature distributions are compared to temperature data obtained from MR thermometry. The study shows that the quality and resolution of the thermometry data is sufficient to validate and improve modeling approaches. Furthermore, the data can be used to identify missing model parameters as well as the exact placement of the laser applicator in relation to the imaging plane.
\end{abstract}

\begin{keywords}
LITT; MR Thermometry; Modeling; Simulation; Experimental Validation
\end{keywords}

\section{Introduction}
\label{sec:introduction}
Laser-induced interstitial thermotherapy (LITT) is a minimally invasive procedure for the local thermal treatment of cancer. In practice, LITT is often used to treat tumors in the liver. For this purpose, a special laser applicator is inserted into the tissue close to the tumor. Laser power is applied directly through the applicator and the tissue around the applicator including the tumor is heated up until coagulative effects cause the destruction of the tumor cells.

One of the main issues of LITT is the planning of the treatment, where the practitioner tries to ensure that all of the tumor cells are being destroyed, while healthy tissue should not be damaged if possible. To aid the practitioners, we consider the simulation of LITT with particular focus on the influence of blood vessels and do a comparison with ex-vivo experiments. 

Models for simulating LITT have been studied for a long time, e.g. \cite{puccini2003simulations, Mohammed, Fasano, Huebner}. The main effects which must be considered in such a model are the propagation of radiation and temperature in the tissue, which are typically modeled by systems of partial differential equations (PDEs). Since the radiation is absorbed by the tissue, it is modeled as a source term for the temperature equation. The local optical properties of the tissue change once coagulation takes place which increases the absorption rate. This effect can be taken into account through a temperature dependent damage parameter which models the transition between the natural and coagulated states. Additionally, the laser applicator includes a water cooling system which is used to postpone the coagulation in proximity to the applicator which would decrease the penetration depth of the radiation.

Two problems arise when dealing with simulations for LITT: First, not all parameters of the model are precisely known or easily measurable and, second, validating the simulation results is no simple task due to the complexity of measuring the temperature and tissue damage during the treatment. In our previous study \cite{Huebner} we used a single temperature probe in order to validate the simulation results. However, comparing the temperatures at a single position may not be enough to validate the simulation and is of course not possible during real treatments. In this paper we want to investigate whether thermometry obtained from MR imaging can be used to validate the simulation results and possibly even help to identify missing parameters. Thermometry uses MR images to derive local temperature measurements. Hence we can not only compare the results of the simulation at a single point, but at the entire image plane of the MR. In particular, we want to investigate whether the accuracy and resolution of the thermometry measurements are sufficient to validate the simulation results and whether this information can be used to identify certain missing parameters for the model.

For this study, four ablation experiments with ex-vivo pig livers were carried out. MR imaging was used throughout all experiments to derive thermometry measurements. Additionally a small tube, acting as an artificial blood vessel, was placed into each of the samples. Water was pumped through the artificial vessel to emulate the cooling effect of blood vessels in the tissue which act as temperature sinks and influence the treatment. The same settings were simulated with our computer model and the temperatures obtained from thermometry and simulation were compared in the two-dimensional imaging plane.

\section{Materials and Methods}
\subsection{Experimental Setup}
Magnetic resonance-guided laser ablation was applied to four porcine liver samples. The liver was obtained from a slaughterhouse and used approximately 6 hours later for the experiments.
For the LITT procedure a ND:YAG laser (MY 30, Martin Medizin-Technik, Tuttlingen, Germany) with a wavelength of 1064 nm and a laser power $\hat{q}_\text{app}$ of 20 W was used. 
A laser applicator (Power‑Laser‑Set; Somatex\SymbReg Medical Technologies, Teltow, Germany) was inserted into the middle of the liver sample.
An optical fiber with a diffuse emission window of 30 mm at the tip delivered the laser radiation into the applicator (Somatex\SymbReg Medical Technologies, Teltow, Germany).
The diffuser tip generates a symmetrical ellipsoidal heating pattern whose longitudinal axis is aligned with the applicator orientation.
The applicator is transparent for the laser radiation and additionally equipped with an internal cooling water circulation system which is used to cool the surface of the applicator and the surrounding tissue.
This is done to prevent overheating and thus carbonization of the tissue as well as damage to the applicator and the optical fiber. The temperature $T_\subcool$ of the cooling water was \SI{20.0}{\celsius}.

\begin{figure}
  \centering
  \includegraphics[width=\textwidth]{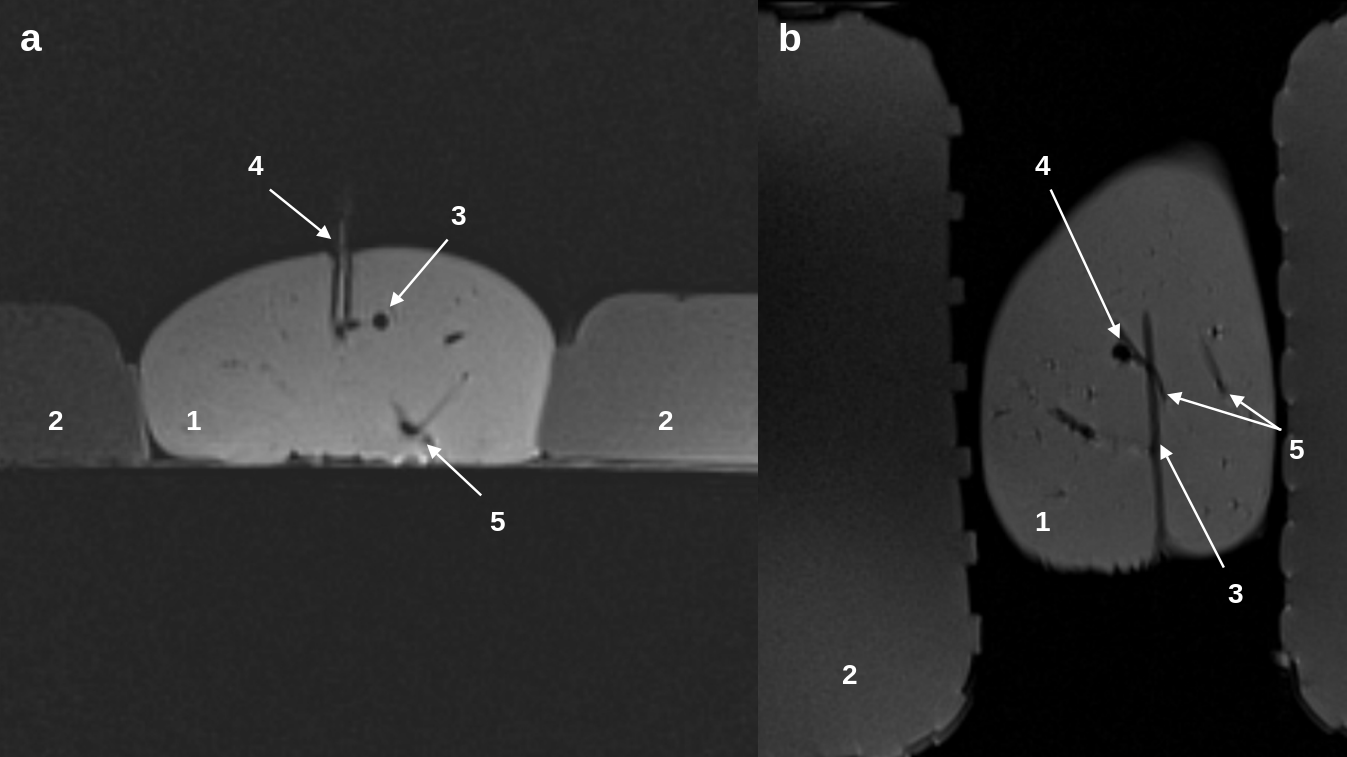}    
	\caption{MR imaging of the experimental setup in axial (a) and coronary (b) orientation. MR images show the liver in the center (1), the agarose gel phantoms in the periphery (2), the laser applicator (3), the plastic tube (4) passing through the liver (partial representation of the plastic tube in a), and bloodless vessels (5).}
  \label{Fig:MR}%
\end{figure}

To emulate blood flow in a vessel in the ablation area, a plastic tube with a diameter of 4 mm was positioned in the liver perpendicular to the applicator shaft (Fig. \ref{Fig:MR}). A roller pump (Dornier Medizintechnik, Wessling, Germany) was used to pass water through the artificial vessel at a flow rate of 60 ml/min. The temperature $T_\subvessel$ of the water was \SI{20.5}{\celsius}.
\begin{table}[htb]
\begin{center}
\begin{tabular}{|l|r|r|r|r|}
\hline
Label & Case 1 & Case 2 & Case 3 & Case 4
\\
\hline
Laser Power $\hat{q}_\text{app}$ [W]  & 
20 & 20 & 20 & 20
\\
\hline
Time Laser on $\laseron$ [s]&
60 & 60 & 60 & 60
\\
\hline
Temperature [$\si{\celsius}$] &&&&\\
\; -initial $T_\subinit$ & 16.0 & 17.5 & 17.5 & 18.0\\
\; -coolant $T_\subcool$ & 20.0 & 20.0 & 20.0 & 20.0\\
\; -vessel $T_\subvessel$ & 20.5 & 20.5 & 20.5 & 20.5\\
\; -ambient $T_\subamb$ & 21.8 & 20.8 & 20.5 & 20.8\\
\hline
\end{tabular}
\end{center}
\caption{Experimental setup for the test cases.}
\label{tbl:testCases}
\end{table}

The positions in the liver were controlled by MR imaging using the 1.5 T scanner Magnetom Aera (Siemens Healthineers, Erlangen, Germany) utilizing a turbo spin echo sequence in coronal and transversal orientation (TR = 700 ms, TE = 12 ms, flip angle = 180, FOV = 220 x 220 \si{\square\milli\meter}, matrix = 256 x 256).
During ablation, a segmented echo planar imaging (seg-EPI) sequence was used for MR thermometry based on the proton resonance frequency shift (PRF) method (TR = 50 ms, TE = 13 ms, flip angle = 12, FOV = 220 x 220 \si{\square\milli\meter}, matrix = 128 x 128, one slice in coronal orientation, TA = 2 s).

The PRF method uses the generated image data of the phase shift to calculate a temperature. The method is described in detail in our previous studies \cite{bazrafshan2019thermometry,bazrafshan2013magnetic,bazrafshan2014temperature}.
Fluoroscopic MR-based thermometry was performed one minute before ablation and for 30 minutes during ablation at 4-second intervals. 
Since a homogeneous magnetic field in the object is essential for the PRF method, two agarose gel phantoms were positioned along the liver sample (Fig. \ref{Fig:MR}) to optimize the adjustments of the MR system (shimming).
The homogeneous phantoms expand the object (liver) for the MR system and thus provide more MR signal to optimize automatic adjustments while reducing susceptibility differences in the peripheral area of the liver.
The gel phantoms with an object size of 220 mm x 90 mm x 35 mm consisted of saline solution (NaCl content: 0.9 \%) and 3 \% agarose.
The initial temperatures $T_\subinit$ of the liver samples were \SI{16.0}{\celsius}, \SI{17.5}{\celsius}, \SI{17.5}{\celsius} and \SI{18.0}{\celsius}. A summary of the setup of the four experiments is given in Table~\ref{tbl:testCases}.

\subsection{Mathematical Model}

\begin{figure}[hbt]
	\centering
	\includegraphics[width=\textwidth]{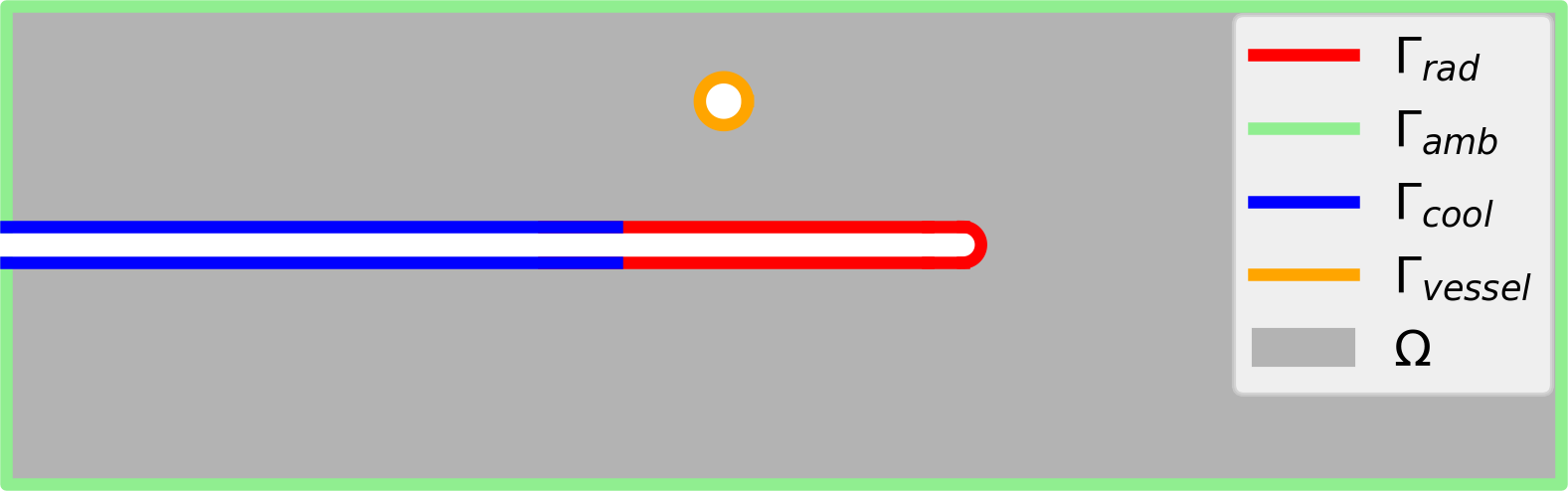}
	\centering
	\caption{Sketch of the computational domain $\Omega$ and the boundary decomposition: radiating surface of the  applicator $\Gamma_\subrad$, cooled surface of the applicator $\Gamma_\subcool$, surface of the artificial blood vessel $\Gamma_\subvessel$, and ambient surface of the liver $\Gamma_\subamb$.}
	\label{fig:sketch_geometry}    
\end{figure}

We denote by $\Omega \subset \R^3$ the computational domain, which is a subset of the liver geometry. Note, that for the computational domain $\Omega$ we do not consider the entire liver, but only a subset given by a sufficiently large box which contains the applicator and the artificial blood vessel. The boundary $\Gamma = \partial\Omega$ consists of the radiating surface of the adjacent applicator $\Gamma_\subrad$, located at the tip of the applicator, the cooled surface of the applicator $\Gamma_\subcool$, the surface at the artificial blood vessel $\Gamma_\subvessel$, and the (artificial) ambient boundary, named $\Gamma_\subamb$ (see Figure~\ref{fig:sketch_geometry}).

The mathematical model is described by a system of partial differential equations (PDEs) for the heat transfer inside the liver, the radiative transfer from the applicator into the liver tissue, and a model for tissue damage (cf. \cite{Mohammed, Fasano, Huebner}), which we explain in the following.  

\paragraph*{Heat Transfer}
The heat transfer in the liver tissue is modeled by the well-known {\it bio-heat equation} (cf. \cite{pennes})
\begin{alignat}{2}
	\label{eq:bioheat}
	\density \heatcapacity \frac{\partial \temperature}{\partial t} - \nabla\cdot\left(\conductivity \nabla \temperature\right) + \perfusion_\subblood(\temperature - \temperature_\subblood) &= Q_\subrad \quad &&\text{ in } (0, \timehorizon) \times \Omega, \\
	\temperature(0,\cdot) &= \temperature_\text{init} \quad &&\text{ in } \Omega, \nonumber
\end{alignat}
where $\temperature = \temperature(x,t)$ denotes the temperature of the tissue, depending on the position $x\in \Omega$ and the time $t\in (0,\timehorizon)$. Here, the end time of the simulation is denoted by $\timehorizon > 0$. Further, $\heatcapacity$ and $\density$ are the specific heat capacity and density of the tissue, respectively, and $\conductivity$ is its thermal conductivity. The perfusion rate due to blood flow is denoted by $\perfusion_\subblood$ and the blood temperature by $\temperature_\subblood$. Note that in the current ex-vivo study the perfusion rate $\perfusion_\subblood$ is set to zero while our artificial blood vessel is modeled through a boundary condition, which is explained below. Finally, $Q_\subrad$ is the energy source term due to the irradiation of the laser fiber defined in \eqref{eq:source_heat} and the initial tissue temperature distribution is given by $\temperature_\text{init}$.

For the heat transfer between the tissue and its surroundings the following boundary conditions are used
\begin{alignat*}{2}
	\conductivity\ \partial_\normal \temperature =&\ \htc_\subcool(\temperature_\subcool - \temperature) \quad &&\text{ on } (0, \timehorizon) \times \left(\Gamma_\subrad\cup \Gamma_\subcool \right), \nonumber\\
	\conductivity\ \partial_\normal \temperature =&\ \htc_\subvessel(\temperature_\subvessel - \temperature) \quad &&\text{ on } (0, \timehorizon) \times \Gamma_\subvessel \\
	\conductivity\ \partial_\normal \temperature =&\ 0 \quad &&\text{ on } (0, \timehorizon) \times \Gamma_\subamb
	.
\end{alignat*}
Here, $\normal$ is the outer unit normal vector on $\Gamma$. Additionally, $\htc_\subcool$ and $\htc_\subvessel$ are the heat transfer coefficients for the water-cooled applicator and the artificial blood vessel, respectively. Here, $\temperature_\subcool$ and $\temperature_\subvessel$ are the temperatures of cooling water through the applicator and the artificial vessel, respectively. Note that $\temperature_\subcool$ and $\temperature_\subvessel$ are assumed to be constant and independent of the flow rates. This simplification is justified as long as the flow rate is high enough such that the increase in coolant temperature is negligible as also discussed in \cite{Huebner}. Furthermore, the temperature flow through the ambient boundary $\Gamma_\subamb$ is assumed to be zero since the ambient boundary is assumed to be far away from the applicator so that there is no heat flux over this boundary.

\paragraph*{Radiative Transfer}

\label{subsec:rte}
In general, the irradiation of laser light is modeled by the {\it radiative transfer equation}
\begin{equation}
	\label{eq:rte}
	s\cdot\nabla I+\left(\absorption{} + \scattering{} \right)I=\frac{\scattering{}}{4\pi}\int\limits_{S^2}P(s\cdot s')I(s',x) \dx{s'} \quad \text{ in } S^2 \times \Omega,
\end{equation}
where the radiative intensity $I = I(s,x)$ depends on a direction $s\in S^2$ on the unit sphere and the position $x\in \Omega$, and $\absorption{}$ and $\scattering{}$ are the absorption and scattering coefficients, respectively. In particular, as that radiative transfer happens significantly faster than temperature transfer, we neglect the time-dependence and use this quasi-stationary model. The scattering phase function $P(s\cdot s')$ is given by the Henyey-Greenstein term which reads (cf. \cite{niemz})
\begin{equation*}
	P(s\cdot s')=\frac{1-\anisotropy{}^2}{(1+\anisotropy{}^2-2\anisotropy{}(s\cdot s'))^{3/2}}\nonumber.
\end{equation*}
Here, $\anisotropy{} \in [-1, 1]$ is the so-called anisotropy factor that describes backward scattering for $g=-1$, isotropic scattering in case $\anisotropy{}=0$ and forward scattering for $\anisotropy{}=1$.

Due to the high dimensionality of the radiative transfer equation \eqref{eq:rte}, we use the so-called $P_1$-approximation to model the radiative energy, the details of which can be found, e.g., in \cite{Modest}. The $P_1$-approximation leads to the much simpler three-dimensional diffusion problem
\begin{alignat}{2}
	\label{eq:p1}
	-\nabla\cdot(\diffusivity\nabla\radiation)+\mu_a \radiation =& 0 \quad &&\text{ in } \Omega,
\end{alignat}
where $\radiation = \radiation(x)$ is the radiative energy and the diffusion coefficient $D$ is given by
\begin{equation*}
	\diffusivity=\frac{1}{3(\absorption{}+(1-\anisotropy{})\scattering{})}.
\end{equation*}

For the radiation equation \eqref{eq:p1}, we use the following set of boundary conditions
\begin{align}
	\label{eq:marshak}
	\begin{aligned}
	\diffusivity \frac{\partial \radiation}{\partial \normal} &= \frac{q_\text{app}}{A_{\Gamma_\subrad}} \quad &&\text{ on } \Gamma_\subrad,
	\\
	\diffusivity \frac{\partial \radiation}{\partial \normal}&=0 \quad &&\text{ on } \Gamma_\subcool, \\
	\diffusivity \frac{\partial \radiation}{\partial \normal}+\frac{1}{2}\radiation&=0 \quad &&\text{ on } \Gamma_\subamb \cup \Gamma_\subvessel,
	\end{aligned}
\end{align}
where $q_\text{app}$ is the laser power entering the tissue and $A_{\Gamma_\subrad}$ the surface area of the radiating part of the applicator. On the ambient and vessel boundaries a Marshak condition is used, see e.g. \cite{Modest}.

We model the radiation entering the tissue in the following way
\begin{equation*}
	q_\text{app} = \begin{cases}
		(1-\coolingfactor{}) \hat{q} & \text{ if } t_\text{on} \leq t, \\
		0 & \text{ otherwise, }
	\end{cases}
\end{equation*}
where $\hat{q}$ is the configured laser power, $t_\text{on}$ is the time, at which the laser is turned on, and the factor $(1-\coolingfactor{})$ models the direct absorption of energy by the coolant (cf. \cite{Huebner}). From the numerical point of view the system given by \eqref{eq:p1} and \eqref{eq:marshak} is much easier to solve than the original system given by \eqref{eq:rte}. Finally, the radiative energy is used to define the source term for the bio-heat equation \eqref{eq:bioheat} in the following way
\begin{equation}
\label{eq:source_heat}
Q_\subrad(x) = \absorption{} \radiation(x).
\end{equation}

\paragraph*{Tissue Damage and Its Influence on Optical Parameters}

The optical parameters $\absorption{}, \scattering{}$ and $\anisotropy{}$ are very sensitive to changes of the tissue's state. In particular, once the coagulation of cells starts, these optical parameters change drastically and, as a result, the radiation cannot enter the tissue as deeply as before. Therefore, we model the damage of the tissue as in, e.g., \cite{Mohammed, Fasano} with the help of the {\it Arrhenius law}, which is given by
\begin{equation}
	\label{eq:arrhenius}
	\damage(t,x)=\int\limits_0^t \frequencyfactor \exp\left(- \frac{\activationenergy}{\gasconstant \temperature(s, x)}\right)\dx{s},
\end{equation}
with so-called frequency factor $A$, activation energy $E_a$, and universal gas constant $R$. This is used to model the change of optical parameters due to coagulation in the following way
\begin{align*}
	\absorption{} =&\ \absorption{\subnative} + (1-e^{-\damage}) (\absorption{\subcoag} - \absorption{\subnative}), \\
	\scattering{} =&\ \scattering{\subnative} + (1-e^{-\damage}) (\scattering{\subcoag} - \scattering{\subnative}), \\
	\anisotropy{} =&\ \anisotropy{\subnative} + (1-e^{-\damage}) (\anisotropy{\subcoag} - \anisotropy{\subnative}),
\end{align*}
where the subscripts $\subnative$ and $\subcoag$ indicate properties of native and coagulated tissue, respectively (cf. \cite{Fasano}). The damage dependence of the empirical absorption factor $\coolingfactor{}$ in the radiation source term is defined in a similar way by
\begin{align*}
\coolingfactor{} =&\ \coolingfactor{\subnative} + (1-e^{-\damage}) (\coolingfactor{\subcoag} - \coolingfactor{\subnative}).
\end{align*}
This is done because coagulation changes the optical properties of the tissue including the refractive index which influences reflection properties at the applicator-tissue-interface. Therefore, the empirical absorption factor should also change between a native $\coolingfactor{\subnative}$ and a coagulated $\coolingfactor{\subcoag}$ state.

\subsection{Numerical Methods}
\label{subsec:numerics}
The mathematical model for radiative heat transfer and the models for vaporization described above were used to simulate the behavior of ex-vivo porcine liver tissue during LITT. 
The computational geometry was generated using Open Cascade (Open Cascade SAS, Guyancourt, France) and the corresponding mesh was created with GMSH, version 2.11.0 (cf. \cite{gmsh}). The governing equations were solved with the finite element method in Python, version 2.7, using the package FEniCS, version 2017.2 (cf. \cite{fenics, fenics_book}). For the numerical solution of the PDEs, we first \mbox{(semi-)discretize} the bio-heat equation in time using the implicit Euler method. Then, we use piecewise linear Lagrange elements for the spatial discretization of the temperature and radiative energy. The resulting sequence of linear systems was then solved with the help of PETSc (cf. \cite{petsc-user-ref}), where we used the conjugate gradient method with a relative tolerance of \num{1e-10}. Afterwards, the damage function is computed element-wise using a right-hand Riemann sum to discretize the time integral of \eqref{eq:arrhenius}. For a more detailed discussion on the numerical simulation we refer to our previous work \cite{Andres2020Identification,Huebner,Blauth2020Mathematical}.

\subsection{Modeling Parameters}
\label{sec:modeling parameters}
A number of parameters are needed to model the evolution of temperature, radiation and damage within the liver tissue. A summary of these is given in Table \ref{table:parameters}. As indicated in the table, most of these parameters were taken from \cite{roggan1995optical, giering1995review, schwarzmaier1998treatment} (cf. \cite{puccini2003simulations}). Moreover, some of the parameters have been fitted for this experimental setting, and the details for the fitting procedure are described below.
\begin{table}[htb]
	\centering
	\begin{tabular}{l r c}
		\toprule
		Parameter & Value & Source \\
		\midrule
		\multicolumn{3}{c}{Optical (native)} \\[0.5em]
		Absorption coefficient $\absorption{\subnative}$ [\si{\per \meter}] & \num{50}
		& \cite{roggan1995optical} \\
		Scattering coefficient $\scattering{\subnative}$ [\si{\per \meter}]  & \num{8000}
		& \cite{roggan1995optical}\\
		Anisotropy factor $\anisotropy{\subnative}$                              & \num{0.97}
		& \cite{roggan1995optical}\\
		Absorption factor $\beta_{qn}$                              & \num{0.14}
		& \cite{Huebner}\\
		\midrule 
		\multicolumn{3}{c}{Optical (coagulated)} \\[0.5em]
		Absorption coefficient $\absorption{\subcoag}$ [\si{\per \meter}]  & \num{60}
		& \cite{roggan1995optical} \\
		Scattering coefficient $\scattering{\subcoag}$ [\si{\per \meter}] & \num{30000}
		& \cite{roggan1995optical} \\
		Anisotropy factor $\anisotropy{\subcoag}$                                & \num{0.95}
		& \cite{roggan1995optical}\\
		Absorption factor $\beta_{qc}$                              & \num{0.35}
		& fitted\\
		\midrule 
		\multicolumn{3}{c}{Thermal} \\[0.5em]
		Thermal conductivity $\conductivity$ [\si{\watt \per \meter \per \kelvin}]    &   \num{0.518} 
		& \cite{giering1995review}\\
		Heat capacity $\heatcapacity$ [\si{\joule \per \kilogram \per \kelvin}]        &  \num{3640} 
		& \cite{giering1995review}\\
		Tissue density $\density$ [\si{\kilogram \per \cubic \meter}]            & \num{1137} 
		& \cite{giering1995review}\\
		Heat transfer coefficient $\htc_\subcool$ [\si{\watt \per \square \meter \per \kelvin}] & \num{250}
		& fitted\\
		Heat transfer coefficient $\htc_\subvessel$ [\si{\watt \per \square \meter \per \kelvin}] & \num{3500} 
		&fitted\\
		\midrule
		\multicolumn{3}{c}{Damage} \\[0.5em]
		Damage rate constant $\frequencyfactor$ [\si{\per \second}]      & \num{3.1e98} 
		& \cite{schwarzmaier1998treatment}\\
		Damage activation energy $\activationenergy$ [\si{\joule \per \mole \per \kelvin}] & \num{6.3e5} 
		& \cite{schwarzmaier1998treatment}\\
		Universal gas constant $\gasconstant$ [\si{\joule \per \mole \per \kelvin}] & \num{8.31}
		& \cite{schwarzmaier1998treatment}\\
		\bottomrule
		\vspace{0.1cm}
	\end{tabular}
	\caption{Modeling parameters for simulating ex-vivo porcine liver tissue with an artificial blood vessel.}
	\label{table:parameters}
\end{table}

\paragraph*{Fitting Missing Parameters} The native coolant absorption factor $\beta_{qn}$ was determined in \cite{Huebner} from the temperature jump which occurs in the coolant temperature at the moment the laser is switched on. But the data obtained from thermometry suggests that this empirical factor does also change when the tissue around the applicator coagulates. Thus, we have introduced an additional coagulated coolant absorption factor $\beta_{qc}$. For now we also treat the heat transfer coefficients $\htc_\subcool$ and $\htc_\subvessel$ as empirical constants even though it would certainly be possible to approximate them using shape and flow characteristics.

This leaves us with three unknown parameters $\beta_{qn}$, $\htc_\subcool$ and $\htc_\subvessel$, which were determined from the data. The least-square distance between thermometry temperature $T_\text{exp}$ and simulated temperature $T_\text{sim}$ was used as objective functions and the L-BFGS-B algorithm \cite{byrd1995limited} from the SciPy optimize package was used to determine the optimal parameters. The resulting parameters were rounded to two significant digits.

\subsection{Computational Domain and Positioning of Applicator and Vessel}

Let $D$ be a box of dimensions 100 x 100 x 100 \si{\cubic\milli\meter}. The computational domain $\Omega \subset D$ was generated by cutting the applicator and vessel from this box. The box was positioned in such a way that the applicator lies in the center of the box. The box is large enough such that effects at the ambient boundary $\Gamma_\subamb$ can be neglected.

The coordinate system for the computational domain is inherited from the MR images such that any image point $(x, y)$ corresponds to the point $(x, y, 0) \in D$. So the MR image is embedded in the plane $\{(x, y, z) \in D; z=0\}$.

The position of the artificial vessel can clearly be seen on the MR images and the corresponding center points $\n{p}_\text{vessel} \in D$ are given in Table \ref{tbl:positions}. The direction of the vessel $\n{d}_\text{vessel} = (0, 0, 1)$ is perpendicular to the image plane. The vessel has a diameter of \SI{4}{\milli\meter} and runs through the entire box $D$.

\begin{table}[htb]
\begin{center}
\resizebox{\textwidth}{!}{\begin{tabular}{l r r r r}
\toprule
Case Label & Case 1 & Case 2 & Case 3 & Case 4 \\
\midrule
Applicator &&&& \\
\; -tip $\n{p}_\text{app}$ [mm] &
(105.6, 120.0, 2.5) & (114.5, 155, 0) & (98.1, 145.3, 1) & (111.3, 131.3, 0)
\\
\; -direction $\n{d}_\text{app}$ &
(0.089, 0.995, 0.042) & (0.033, 0.999, 0) & (-0.115, 0.993, 0.003) & (-0.080, 0.997, 0)
\\
\midrule
Vessel &&&& \\
\; -point $\n{p}_\text{vessel}$ [mm] &
(109.3, 86.8, 0) & (122.6, 124.9, 0) & (109.5, 132.7, 0) & (105.3, 114.8, 0)
\\
\; -direction $\n{d}_\text{vessel}$ &
(0, 0, 1) & (0, 0, 1) & (0, 0, 1) & (0, 0, 1)\\
\bottomrule
\end{tabular}}
\end{center}
\caption{Positions of the Applicator and the Artificial Vessel}
\label{tbl:positions}
\end{table}

In the experiments the applicator was aligned with the image plane as closely as possible. However, there is still a slight deviation which can have a significant effect when comparing the measured and simulated temperatures. Therefore, we have corrected the applicator tip position $\n{p}_\text{app}$ and direction $\n{d}_\text{app}$ using a parameter fitting with the least-square distance of the temperatures as an objective, analogously to the fitting procedure described previously. The corrected positions which were used for the simulation are given in Table \ref{tbl:positions}.

\section{Results}
\begin{figure}[p]
\includegraphics[width=\textwidth]{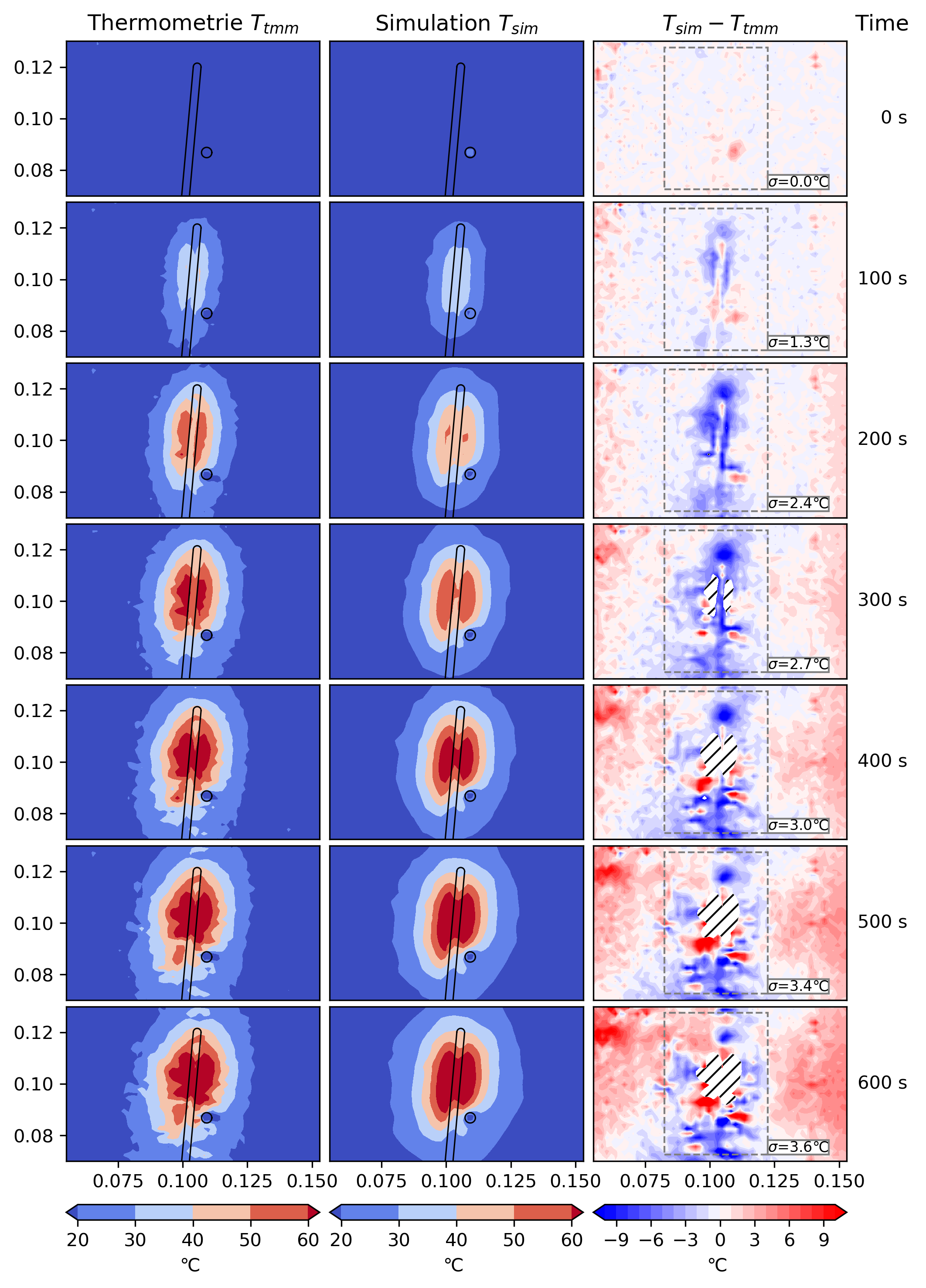}
\caption{Case 1: Comparison between measured $T_\text{exp}$ and simulated temperature $T_\text{sim}$. The standard deviation $\sigma$ is computed within the dashed box, disregarding the hatched area where the measurement is unreliable due to coagulation.}
\label{fig:case1}
\end{figure}
\begin{figure}[p]
\includegraphics[width=\textwidth]{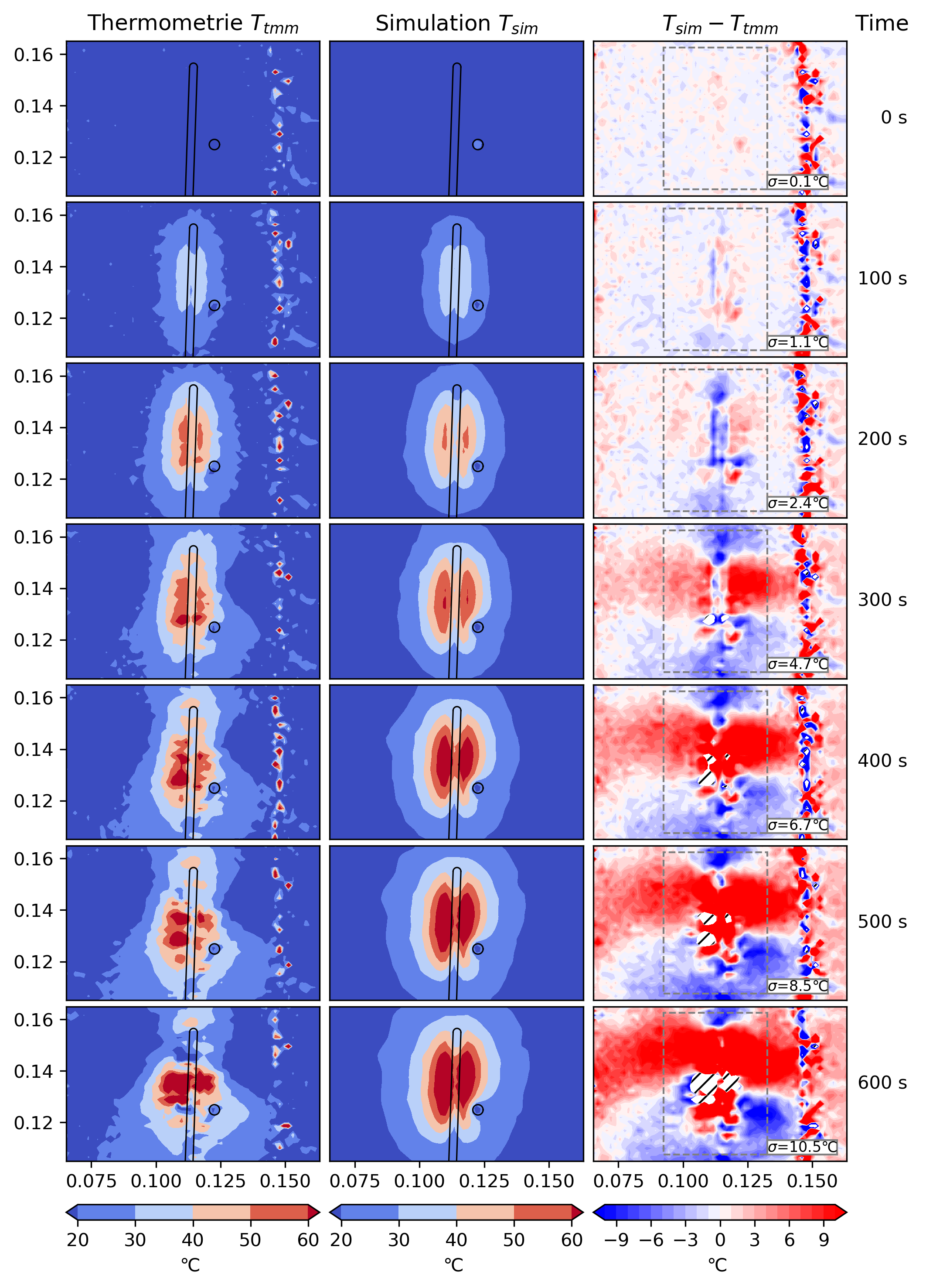}
\caption{Case 2: Comparison between measured $T_\text{exp}$ and simulated temperature $T_\text{sim}$. The standard deviation $\sigma$ is computed within the dashed box, disregarding the hatched area where the measurement is unreliable due to coagulation. \textbf{Note:} For this case artifacts in the MR thermometry data likely result in faulty temperature measurements.}
\label{fig:case2}
\end{figure}
\begin{figure}[p]
\includegraphics[width=\textwidth]{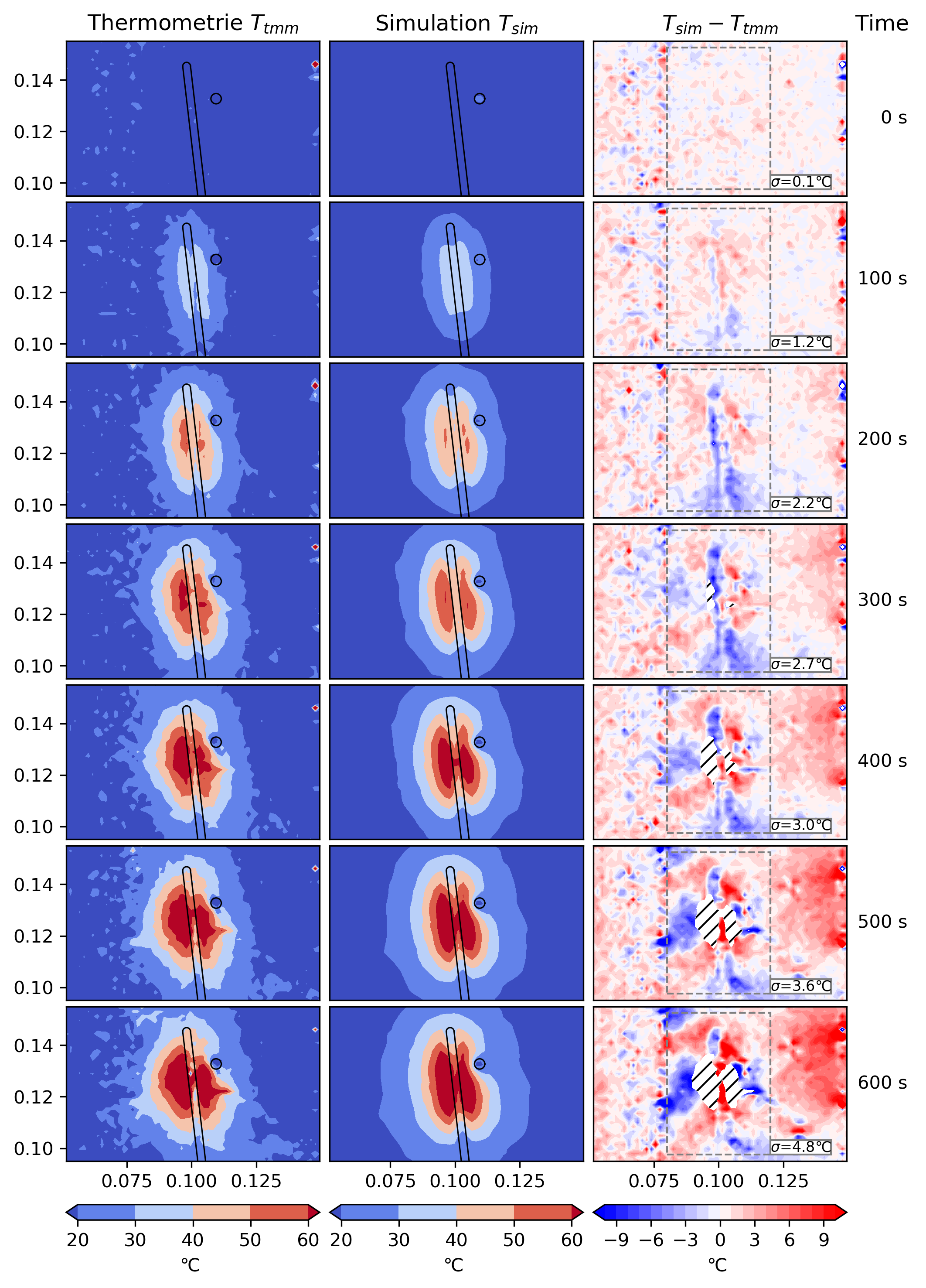}
\caption{Case 3: Comparison between measured $T_\text{exp}$ and simulated temperature $T_\text{sim}$. The standard deviation $\sigma$ is computed within the dashed box, disregarding the hatched area where the measurement is unreliable due to coagulation.}
\label{fig:case3}
\end{figure}
\begin{figure}[p]
\includegraphics[width=\textwidth]{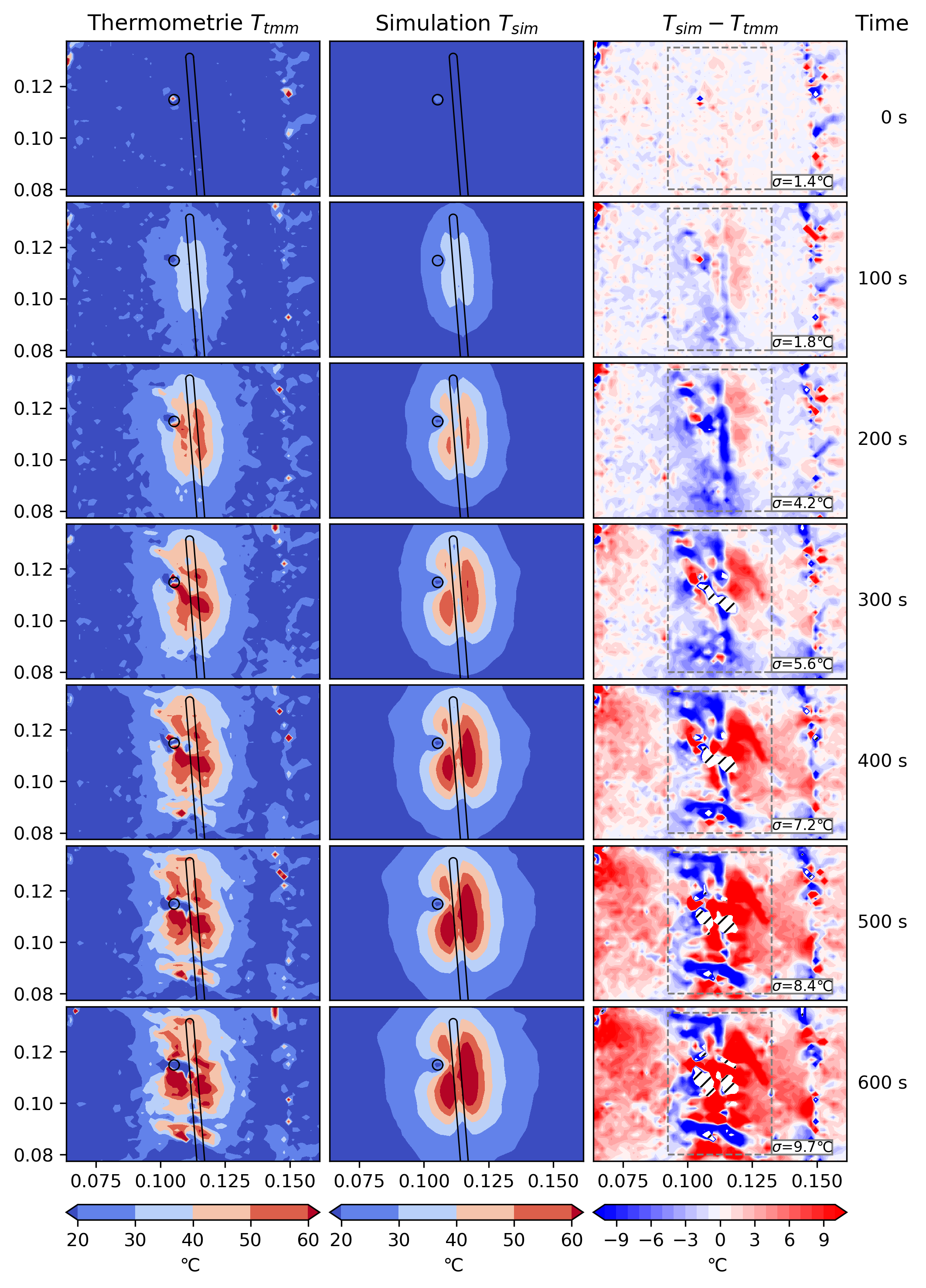}
\caption{Case 4: Comparison between measured $T_\text{exp}$ and simulated temperature $T_\text{sim}$. The standard deviation $\sigma$ is computed within the dashed box, disregarding the hatched area where the measurement is unreliable due to coagulation. \textbf{Note:} For this case artifacts in the MR thermometry data likely result in faulty temperature measurements.}
\label{fig:case4}
\end{figure}
The results for the four experiments Case 1-4 are shown in Figs. \ref{fig:case1}-\ref{fig:case4}, respectively. Each figure shows a comparison between the measured temperature $T_\text{exp}$ obtained from MR thermometry in the left column and the simulated temperature $T_\text{sim}$ in the middle column over a time range between \SI{0}{\second} to \SI{600}{\second} into the experiment. The positions of the applicator and the artificial vessel are marked.

The right column shows the difference between simulated and measured temperature. The measured temperature $T_\text{exp}$ is not reliable after coagulation takes place due to the corresponding changes of the material parameters, which makes the thermometry data invalid. In the right column of Figs. \ref{fig:case1}-\ref{fig:case4} this is indicated by a hatched region
\begin{align*}
\mathcal{S}_\text{hatch}(t) = \{(x, y, 0) \in \Omega| T_\text{exp} > \SI{60}{\celsius}\},
\end{align*}
where the measured temperature $T_\text{exp}$ and, thus, also the difference $T_\text{sim}-T_\text{exp}$ is not meaningful anymore. The deviation between simulated and measured temperature is quantified by the standard deviation $\sigma$ of $T_\text{sim}-T_\text{exp}$ printed in the right column. In order to exclude noisy thermometry data, the standard deviation is computed within the active region of the ablation which is marked by a dashed gray line. Values within $\mathcal{S}_\text{hatch}(t)$ where the temperature difference is not meaningful are ignored when computing the standard deviation.

\section{Discussion}
The agreement between measured and simulated temperature is very good for Case 1 (Fig. \ref{fig:case1}) and Case 3 (Fig. \ref{fig:case3}), where we have a standard deviation $\sigma$ of \SI{3.6}{\celsius} and \SI{4.8}{\celsius}, respectively, after \SI{600}{\second}. Additionally, the contour lines of the temperatures are very similar. The effect of the artificial vessel is clearly visible and matches between measurement and simulation. The measured data shows minor artifacts which do not cause a problem for the comparison.

For Case 2 (Fig. \ref{fig:case2}) and Case 4 (Fig. \ref{fig:case4}) there seems to be a major problem with the measured data. Fig. \ref{fig:case2} shows a temperature drift in a wide horizontal strip in the middle of the observation area. In part, the measured temperature actually decreases, even in distant areas on which the ablation has no influence at all. A similar effect can be seen in Fig. \ref{fig:case4}. As a consequence, also the computed standard deviations are significantly larger, with a value of \SI{10.5}{\celsius} and \SI{9.7}{\celsius}, respectively, after \SI{600}{\second}.

The deviations encountered in Cases 2 and 4 likely originate from measurement errors: Such artifacts can result from changes of the magnetic field in the PRF method. A drift of the static magnetic field (B0 drift) can occur at long acquisition times by the scanner system but its effects influence the whole image. Also susceptibility differences between liver tissue and air may have produced field inhomogeneities during the treatment \cite{kuroda2018mr}. Such inhomogeneities even more occur with higher temperatures ($>\SI{100}{\celsius}$) and the associated gas evolution in the liver. The gases can disperse in the empty vessels of the ex-vivo liver and lead to incorrect phase values for temperature calculation \cite{kuroda2018mr}. This issue mainly exists for ex-vivo experiments where the liver is not perfused.

The following measures were taken to reduce artifacts: The liver was framed by two agarose gel phantoms (with similar susceptibility as the liver tissue) in the experiments to significantly increase the object volume and thus help stabilize the magnetic field \cite{saccomandi2013techniques, hubner2013influence} and it was also tried to correct for changes in the magnetic field (cf. \cite{poorman2019multi, odeen2019magnetic, kagebein2018motion}).

\section{Conclusion}

In this paper, we have investigated the validation of ex-vivo laser induced thermotherapy (LITT) simulations with the help of MR-thermometry data. We presented a mathematical simulation model for LITT and described the experimental setup used to validate our model. Four experiments were carried out, and the temperature measurements obtained from thermometry were found to be very suitable for validating simulation results and identifying missing parameters. Apart from the likely measurement errors in Cases 2 and 4, the data is of high quality and plausible. The resolution of the data is sufficiently high to resolve even small details such as the influence of the artificial blood vessel on the temperature distribution. Simulation results were in good agreement with measurements and it was possible to identify missing parameters from the data. Altogether, thermometry is a powerful tool to validate and improve simulation models, in particular, since measurements are not restricted to pre-selected points, but are available throughout the entire imaging plane.

\section*{Funding}
This work was supported by the Federal Ministry of Education and Research of Germany (BMBF) under Grant 05M16AMA. 

\section*{Competing interests}
  The authors report no conflicts of interest.

\bibliographystyle{plain}
\bibliography{references.bib}

\end{document}